\documentstyle[12pt,epsf]{article}
\begin{document}
\pagestyle{plain}
\setcounter{page}{1}
\subsection*{\centerline{An alternative interpretation of GRB 970228}}
\subsection*{\centerline{data and nature}}
\vskip 1 cm
\centerline{\bf{D.Fargion $^{1,2,3}$, A.Salis $^1$}}
\centerline{}
\centerline{$^1$ Dipartimento di Fisica, Universita' degli Studi di Roma}
\centerline{"La Sapienza" P.le A. Moro 2, 00185 Rome, Italy. }
\centerline{$^2$ INFN, RomaI, Italy}
\centerline{$^3$ Technion Institute, Haifa, Israel}
\vskip 1cm
\begin{abstract}
Time evolution and signature of GRB 970228 are analysed in the framework of 
fireball or precessing gamma jet models
\end{abstract}
\vskip 1cm

\centerline{\bf{Fireball versus Precessing Gamma Jet}}
\vskip 0.2cm

The very last news of a discovery of a fading X burst associated to GRB 970228 
[1,2] by BeppoSAX experiment as well as the early identification of an optical 
counterpart had led to the prompt claim [3,4] of an extragalactic nature of 
GRBs. Most cosmological GRB models are associated to huge energetic 
($E_{GRB}\sim 10^{51}~erg$) one-shot spherically symmetric fireball event. The 
discovery overshadowed a previous convincing sequence of GRBs found by BATSE 
from the same sky direction [5] and whose natural interpretation strongly 
favoured a repeater nature of GRB (see also [6]) and consequently a precessing 
(and blazing) Gamma Jet Model [7,8,9]. Successive (1) optical observations by 
Earth telescopes [3,4] 21 hours after the GRB 970228 [10] at magnitude $m_I=
20.6$, (2) the dimming and disappearance of the source, (3) later optical 
identification at the same 
position, at lower intensity (23.5 magnitudes), of an extended relic image (a 
galaxy?) on 13 March, led (4) to the claim of an extragalactic nature of the 
event [3,4]. However the same fading and disappearance of that diffused 
(galactic?) image (since 13 March till 26 March) make unprobable or at least 
premature (at that stage) 
any clear identification with (stable) galaxy source. Within the fireball 
model these images evolution might be associated only to the fading tail of the 
explosion. Moreover, if the fireball intensity since 13 March is comparable 
(or hidden) in the 23.5 magnitudes galactic foreground, the question is why 
the last 26 March image by HST [3] at 25.7 magnitudes exhibits a clear higher 
intensity ratio of the large (pointlike) spot over the diffused and extended 
(galactic?) background. Any deformation or fading of the diffused (extended) 
source in any near future observation might probe a different nature of the 
event: for instance a relic cooling tail of a jet. Analogous images, but at 
larger scales, are already observed in Seyfert or Quasars like 3C273. The fine 
structure variability of GRBs ($\ll~sec$) is not 
compatible to Seyfert, QSO objects and corresponding Schwartzchild times (even 
at reasonable superluminal regime). Therefore only small precessing 
gamma jets (NS, BH of solar masses) might explain [5,6,7] the event. Moreover 
the possible 
identification of the extended source with the asymmetric jet tail size 
(nearly 1/4 arcsec) and the corresponding (one month) scale time define 
naturally a GRB distance within galactic halo models. Near future optical 
observations (few weeks) may better clarify this hot issue.

\centerline{\bf{References}}
\vskip 0.2cm

[1] Costa E. et al., 1997a, IAU Circular 6572
\newline
[2] Costa E. et al., 1997b, IAU Circular 6576
\newline
[3] NASA MSFC Astronomy headlines 3/3/97;\newline
NASA Press Release STScI-PR97-10 1/4/97
\newline
[4] Van Paradijs J. et al., Nature {\bf 386} 686, (1997)
\newline
[5] NASA MSFC Astronomy headlines 17/12/96
\newline
[6] Quashnock J.M., Lamb D.Q., MNRAS {\bf 265} 159, (1993)
\newline
[7] Fargion D., Salis A., Nucl.Phys.B (Proc.Suppl.) {\bf 43} 269, (1995)
\newline
[8] Fargion D., Salis A., ApSS {\bf 231} 191, (1995)
\newline
[9] Blackman E.G., Yi I., Field G.B., Ap.J.Lett. in press 1997
\newline
[10] Grott P.J. et al. 1997, IAU Circular 6584

\end{document}